\documentclass[twocolumn,twoside,slac_two]{revtex4}
\usepackage{graphicx}
\usepackage{fancyhdr}
\begin{document}

\title{Inclusive semileptonic $B$ decays: $|V_{cb}|$ and $|V_{ub}|$}


\author{M. Rotondo}
\affiliation{INFN sezione di Padova, Italy}

\begin{abstract}
The present status of the measurement of the inclusive semileptonic $B$ decays 
is reviewed. In particular the determination of the Cabibbo-Kobayashi-Maskawa 
matrix elements $|V_{cb}|$ and $|V_{ub }|$ is discussed and some future prospects 
are given. 
\end{abstract}

\maketitle

\thispagestyle{fancy}

\section{Introduction}
With the discovery of CP violation in the $B^0$ mesons, and the precise measurement 
of the angle $\beta$ \cite{beta} of the Unitarity Triangle (UT), the experimental 
effort focus on the measurements of other parameters to over-constraint the 
Cabibbo-Kobayashi-Maskawa (CKM) matrix \cite{CKM}.  
In particular the information on the side $R_b$ opposite to the angle $\beta$ is crucial
to test the standard model prediction for CP violation. 
In the Wolfenstain parametrization $|R_b|=\lambda^{-1}(1-\lambda^2/2)|V_{ub}/V_{cb}|$
where $\lambda=|V_{us}|/\sqrt{|V_{us}|^2+|V_{ud}|^2}\approx 0.226$, so a precision
measurement of $R_b$ require the measurement of $|V_{ub}|$ and $|V_{cb}|$ with high accuracy. 
Moreover the parameters $|V_{ub}|$ and $|V_{cb}|$ play a special role in
the CKM matrix, because they can be extracted from pure tree level decays, so
their values are to high accuracy, independent of any new physics contributions.

%The parameter $|V_{cb}|$ is known, from a global fit to the heavy quark parameter, 
%with an uncertainties of about $2\%$, so at present the 
%uncertainty on $|V_{ub}|$, close to $9\%$, is the limiting factor to the constraint. 
%Current world averages are provided by the Heavy Flavour Averaging Group (HFAG) \cite{HFAG}.

The semileptonic transitions $b\to c\ell\nu$ and $b\to u\ell\nu$ are characterized by an 
electroweak current that probes the 
$B$ dynamics allowing to determine the CKM matrix element $|V_{xb}|$ in a clear 
environment. Their simplicity is only apparent, because we are interested in precision measurements, 
the complexity specific of QCD dynamics have to be addressed and taken in consideration.
Both inclusive and exclusive final states can be exploited to extract the CKM matrix
elements. The experimental and theoretical techniques underlying these two 
approaches are different and provide crucial cross-check on our understanding
of the QCD frameworks. At present the inclusive determinations for both $|V_{cb}|$
and $|V_{ub}|$ are more precise than the exclusive determinations.
%$|V_{cb}|$ is known, from inclusive determinations, 
%with an uncertainties of about $2\%$, $|V_{ub}|$ with un uncertainty close to $9\%$.  
However improvements of the  exclusive determinations,
whose uncertainties are dominated by the lattice QCD 
calculations, are an important goal for the next future. 

A comprehensive review of the semileptonic measurements is beyond the scope of
this paper. We will present only the most recent measurements of both $|V_{ub}|$ 
and $|V_{cb}|$. 

\section{Inclusive $B\to X_c\ell\nu$ decays}

The theoretical approaches used to extract both $|V_{cb}|$ and  $|V_{ub}|$ from inclusive
$B\to X\ell\nu$ decays use the fact that the $b$-quark mass is large compared to the 
$\Lambda_{QCD}$ scale, that caracterize the low energy hadronic physics.
This allows to use an effective-field-theory, the Heavy Quark Expansion (HQE), to 
separate the non-perturbative from perturbative contributions, and write a
double expansion in powers of $\Lambda_{QCD}/m_b$ and $\alpha_s(\mu)$, 
with $\mu>>\Lambda_{QCD}$. 
In this framework the total semileptonic $B\to X_c\ell\nu$ decay widht is given by 
\begin{equation}
\Gamma_{sl}=|V_{cb}|^2 \frac{G^2_F m_b^5}{192\pi^3}(1+A_{ew})A^{pert} F(r,\frac{\mu_\pi^{2}}{m_b^2},\frac{\mu_G^{2}}{m_b^2},...)\\
\label{eq:gammasl}
\end{equation}
where $r=m_c/m_b$, $A_{ew}$ are the electrowek corrections, $A^{pert}$ summarize 
the QCD radiative corrections,
and the term $F$ is written as an espansion in powers of $1/m_b$, and depends on 
non-perturbative parameters. At each order in $1/m_b$ new non-perturbative parameters come out. At 
present this espansion is computer till the $1/m_b^4$ term.  
A crucial feature of the $F$ espansion is that the first order term is zero, 
making the leading term in this expansion precise at the $\%$ level, and the higher order just small 
corrections to the leading term.

Expressions similar to Eq.\ref{eq:gammasl} can be computed for the moments of 
the lepton momentum $p_\ell$ spectra, and the moments of the squared hadronic mass spectra $m_X^2$ in 
$B\to X_c\ell\nu$ decays, together with the moments of the energy of the $\gamma$ emitted
in the inclusive radiative $B\to X_s\gamma$ decays. This allows to combine the measured
$\Gamma_{sl}$ with the measurements of the moments of other kinematic variables to
determine $|V_{cb}|$ with high precision, togheter with non-perturbative HQE parameters
and $b$- and $c$-quark masses. 
   
Measurements of hadronic mass distribution and leptonic spectrum 
have been made by many experiments CLEO\cite{CLEO_mom}, {\sc BaBar} \cite{BaBar_mom}, 
Belle \cite{Belle_mom},  DELPHI \cite{DELPHI_mom} and CDF \cite{CDF_mom} (the latter 
provides only the measurement of the hadronic moments).
 
At the modern $B$-factories, the large samples of $\Upsilon(4S)\to B\bar B$, 
can be exploited to provide a clean sample of tagged $B$ meson, by reconstructing one of the 
$B$ meson ($B_{reco}$) via fully hadronic modes, $B_{reco}\to D^{(*)}Y^{\pm}$, where $Y^{\pm}$
corresponds to a combination of $\pi^{0/+}$'s and $K^{0/+}$'s. 
The kinematic consistency of the $B_{reco}$ with a $B$ meson is tested using the variables 
$m_{ES}=\sqrt{s/4-{\bf p}_B^2}$ and $\Delta E=E_B-\sqrt{s}/2$, where 
$\sqrt{s}$ is the center of mass energy, and $E_{B}$ and ${\bf p}_B$ are the energy and the momentum
of the $B_{reco}$ candidate in the $\Upsilon(4S)$ frame. 
By fully reconstructing one of the $B$ mesons in the events, the charge, flavor and momentum of 
the second $B$ can be inferred. By knowing event by event the momentum of the 
signal $B$ meson, the kinematics of the decay product of the signal $B$ meson be computed 
in the center of mass of the $B$ meson, reducing the uncertainty due to the motion of the $B$
meson in the $\Upsilon(4S)$ frame. 

Both {\sc BaBar} and Belle have used the $B_{reco}$ sample for many studies on the $B$ semileptonic 
decays. The efficiency to reconstruct a $B_{reco}$ candidate is quite low, $0.3\%$ for $B^0{\overline B^0}$ 
and $0.5\%$ for $B^+B^-$ events, but the purity of the sample is higher than $80\%$ allowing to 
reduce systematic uncertainties due to backgrounds knowledge.  

{\sc BaBar} recently presents \cite{BaBar_newmom} a measurement of the moments of the hadronic 
mass $\langle m_X^k \rangle$, with $k=1...6$,
using a larger data-set ($220\cdot 10^6$ $B\overline B$) than the previous measurement 
(based on $88\cdot 10^6$ $B\overline B$) \cite{BaBar_mom}. 
Moreover, {\sc BaBar} present for the first time the measurement of the mixed hadronic 
mass-energy moments $\langle n_X^k \rangle$, with $k=2,4,6$, where 
$n_x$ is defined by $n_x^2=m_x^2-2\Lambda E_X+\Lambda^2$, 
$m_X$ and $E_X$ are respectively the mass and the
energy of the $X_c$ system in the $B$ rest frame, and $\Lambda=0.65$~GeV. 
The $n_X$ moments has been proposed in Ref.\cite{gambino_mixmom} due to their high 
sensitivity to higher order non-perturbative parameters. 

\begin{figure}[t]
\centering
\includegraphics[width=80mm]{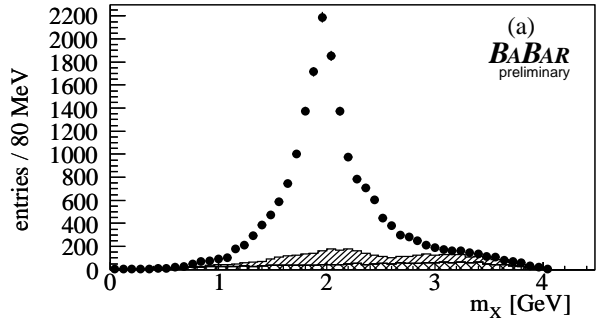}
\caption{
The measured hadronic mass spectrum for $p_{\ell,min}>0.8$~GeV in the $B$ rest frame.
The tag-side background (hatched histogram) and the signal side background (cross-hatched histogram)
are superimposed to the data.} \label{f:babar_mxhad}
\end{figure}

The analysis uses the $B_{reco}$ sample, and proceed reconstructing a $B_{reco}$, 
and identifying a lepton in the event (both electron and muons are used). 
All particles, both neutral and charged, that are not used in the reconstruction of 
the $B_{reco}$ are assigned to the signal $B$. A kinematical fit imposes 
4-momentum conservation, the equality of the masses of the two $B$ mesons, and 
constrains the mass of the neutrino, inferred from the missing momentum $P_{\nu}$, 
to zero with $P_\nu^2=0$. 
This allows to reconstruct with a good resolution the distribution of 
the mass of the hadronic system, $m_X$. An example of
the hadronic mass spectra is shown in Fig.\ref{f:babar_mxhad}. To correct for the
effect of the lost particles, {\sc BaBar} measures the hadronic moments using 
calibration curves obtained with the Monte Carlo that relate the measured moments 
to the true moments.
\begin{figure*}[t]
\centering
\includegraphics[width=160mm]{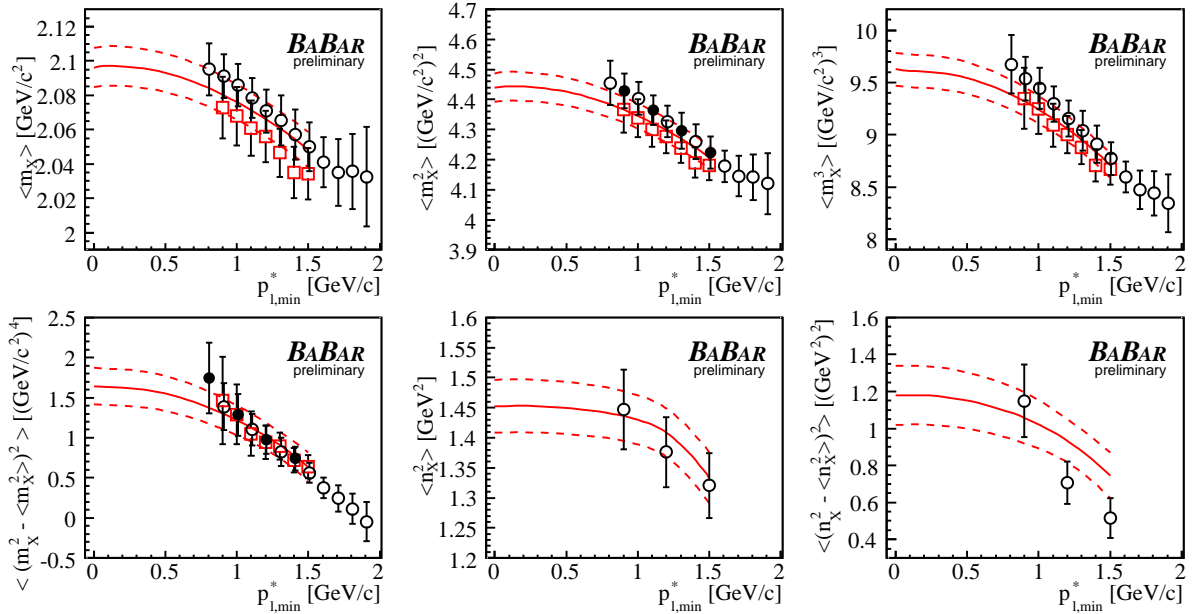}
\caption{The measured hadronic mass and mixed moments, as a function of the minimum lepton
momentum $p_{\ell,min}$ compared with the result of the HQE fit (solid line). The previous 
measurement by {\sc BaBar}, red square, are also superimposed. Only the solid circles are
included in the fit. The dashed line correspond to the fit uncertainty propagated to the individual moments.
The measured moments continue to decrease increasing $p_{\ell,min}$, and extend beyond theoretical 
predictions that are available for $p_{\ell,min}<1.5$~GeV/$c$. } 
\label{f:babar_hadmom}
\end{figure*}
The main sources of systematics errors on the moments are due to the uncertainties related to
the detector efficiency for tracks and photons, and to the $X_c$ signal modelling, mainly due to
the poor knowledge of the $B\to D^{**}\ell\nu$ composition, and particular care is also 
devoted to take out the radiative correction.

{\sc BaBar} measures the hadronic and mixed moments as a function of the lower
cut on the lepton momentum $p_{\ell,min}$ between  $p_\ell>0.8$ to $p_\ell>1.9$ GeV/$c$
in the $B$ rest frame. The results are compatible with the previous measurement \cite{BaBar_mom}, 
a comparison is shown in Fig.\ref{f:babar_hadmom} togheter with the  results of the HQE fit 
described in the next section.
The very high lepton momentum cuts, even if not used to constrain the HQE parameters, 
is important to test the Operator Product Expansion (OPE) in a region dominated 
only by the $D$ and the $D^*$ exclusive states. 

\subsection{Determination of HQE parameters and $|V_{cb}|$}

{\sc BaBar} performs a combined fit to the hadronic mass moments measurement presented above, the 
lepton energy moments in $B\to X_c\ell\nu$ decays \cite{BaBar_mom} and photon energy moments 
in $B\to X_s\gamma$ decays \cite{BaBar_gamma}, in the framework of the kinetic scheme \cite{kinetic}. 
The results (we do not report the higher order HQE parameters that are left free in the fit) are 
\begin{eqnarray}
|V_{cb}|\cdot 10^{3}  & = &  41.88\pm 0.44_{exp}\pm 0.35_{theo}\pm 0.59_{\Gamma_{SL}} \nonumber \\ 
m_b      & = & (4.552\pm 0.038_{exp}0.040\pm_{theo}) {\rm GeV}\nonumber \\
\mu_\pi^2& = &(0.471\pm 0.034_{exp}0.062_{theo}) {\rm GeV}^2 \nonumber,
\label{babar_ris}
\end{eqnarray}
\noindent where the additional uncertainty of $1.4\%$ for the $|V_{cb}|$ is a normalization 
uncertainty due to non calculated terms in the total rate. The quality of the fit is very good, 
with a $\chi^2=8$ for 20 degree of freedom. 
In Fig.\ref{f:babar_hadmom} the fit results, using only the {\sc BaBar} data, 
is superimposed to the measured moments. The moments $\langle m_X\rangle$ and 
$\langle m_X^3\rangle$ are not included in the fit but provide an unbiased 
comparison with the fitted HQE prediction. 
The measured moments $\langle n_X^2\rangle$ and 
$\langle (n_X^2 - \langle n_X\rangle^2)^2\rangle$, also are not included in the fit, 
but are in good agreement with the prediction.

Recently {Belle} re-analyze the $B\to X_s\gamma$ data published 
in Ref.\cite{Belle_egamma} and measures the first and the second moment of the photon
energy spectrum with various cuts on the minimum energy threshold, $E_{min}=1.9,... 2.3$~GeV.
The new measured moments \cite{Belle_newegamma} are in agreement with the 
{\sc BaBar}\cite{BaBar_gamma} and CLEO \cite{CLEO_egamma} measurements. 
Using these new moments, Belle performs a combined fit with 
the recent Bella data on the lepton energy and hadronic 
mass moments \cite{Belle_mom} in $B\to X_c\ell\nu$ decays.  The fit has been performed 
in both the kinetic scheme  
and the 1S scheme \cite{1s}. The results are reported in Tab.\ref{t:belle_global} together
with the result using {\sc BaBar} only moments, and the results from the global HQE 
fit including all the moment measurements from {\sc BaBar}, Belle, CDF, CLEO and DELPHI
(see HFAG web page for more details \cite{HFAG}). 
\begin{table}[h]
\begin{center}
\caption{ Fitted values for $|V_{cb}|$, $m_b$ and the $\mu_\pi^2$ ($\lambda_1$) parameters
for the new Belle and {\sc BaBar} fits, described in the text. For comparison the results of the global fit
provided by HFAG are also reported. {\sc BaBar} reported results in the kinetic scheme, 
Belle reported results also in the 1S scheme.}
\begin{tabular}{l|c|c|c}
\hline \textbf{Kinetic scheme} & \textbf{$|V_{cb}|\times 10^3$}    & $m_b^{kin}$           & $\mu_{\pi}^2$~GeV$^2$\\
\hline 
Belle                          & 41.52$\pm$0.90    & 4.543$\pm$0.075      & 0.539$\pm$0.079 \\
Belle only $X_c\ell\nu$        & 41.46$\pm$0.99    & 4.573$\pm$0.134      & 0.523$\pm$0.106 \\
\hline
Babar                          & 41.88$\pm$0.81    & 4.552$\pm$0.055      & 0.471$\pm$0.070 \\
\hline
Global                         & 41.91$\pm$0.68    & 4.573$\pm$0.034      & 0.408$\pm$0.035 \\
Global only $X_c\ell\nu$       & 41.68$\pm$0.70    & 4.677$\pm$0.053      & 0.387$\pm$0.039 \\
\hline
\end{tabular}
\begin{tabular}{l|c|c|c}
\hline \textbf{1S scheme} & \textbf{$|V_{cb}|\times 10^3$}    & $m_b^{1S}$ & $\lambda_1$ ~GeV$^2$\\
\hline 
Belle                               & 41.56$\pm$0.68    & 4.723$\pm$0.055      & -0.303$\pm$0.046 \\
Belle only $X_c\ell\nu$             & 41.55$\pm$0.80    & 4.718$\pm$0.119      & -0.308$\pm$0.092 \\
\hline
Global                               & 41.78$\pm$0.31    & 4.701$\pm$0.030      & -0.313$\pm$0.025 \\
Global only $X_c\ell\nu$             & 41.56$\pm$0.40    & 4.718$\pm$0.058      & -0.274$\pm$0.047 \\
\hline
\end{tabular}
\label{t:belle_global}
\end{center}
\end{table}
It can be seen
that including the photon energy moments reduces substantially the uncertainty on the 
$b$-quark mass and the $\mu_\pi^2$ parameter. In the kinetic scheme, the error on $m_b$ 
is only $34$~MeV, this is crucial to reduce the uncertainty for the $|V_{ub}|$. 
But it has been recently argued that the $B\to X_s\gamma$ input should not be used to
constrain the HQE parameters, due to some model dependence \cite{Neubert_lp07}. 
However the {\sc BaBar} and Belle only fits, and also the global fits, show good agreements, within the 
present uncertainty, between the fit with and without the inclusion of the 
$B\to X_s\gamma$, see Tab.\ref{t:belle_global}.
The $\chi^2/{\rm d.o.f.}$ are very good, and are always well below 1, which could be an hint that 
theoretical errors are overestimated, or theoretical correlations are not 
correctly accounted in the fits. Further investigation are needed in the next future. 
The fits in the $1S$ and the kinetic scheme agree very well, and the uncertainty on $|V_{cb}|$ are 
below $2\%$. The $b-$quark mass, $m_b$, and the HQE parameters must be translated in the same scheme 
to be compared. After  the translation the agreement is quite good if the uncertainties 
due to the scheme translation are properly included. 

Some improvements can be expected in the future with the inclusion in the global fit of the 
higher order moments measured by {\sc BaBar}, that may improve the determination of the 
higher order non-perturbative HQE parameters. 

It should here be reported, that the exclusive determination of $|V_{cb}|$ using the recent $B\to D^*\ell\nu$ 
form factor computation \cite{laiho}, differs by the inclusive determination by more than $2\sigma$. 
Further studies are needed on both theoretical and experimantal side.
  
%
%\begin{figure}[h]
%%\includegraphics[width=60mm]{kin_mbvcb.eps} \\
%\includegraphics[width=60mm]{1s_mbvcb.eps} 
%\caption{Example of a One-column Figure.} \label{example_figure}
%\end{figure}
%\begin{figure}[h]
%\begin{tabular}{c}
%\centering
%\includegraphics[width=40mm]{kin_mbvcb.eps} \\ %&\includegraphics[width=40mm]{1s_mbvcb.eps} \\
%\caption{Example of a One-column Figure.} \label{example_figure}
%\end{tabular}
%\end{figure}

\section{Inclusive $B\to X_u\ell\nu$ decays}

As for the $B\to X_c \ell\nu$ decays, see Eq.\ref{eq:gammasl}, the full rate for $B\to X_u \ell\nu$ decays 
is proportional to $|V_{ub}|\cdot m_b^5$, times a function that accounts for 
QCD and electroweak corrections. The theoretical framework also in this case, 
is the HQE, which predict the total $\Gamma(B\to X_u\ell\nu)$ decay rate
with uncertainty of about $5\%$ \cite{opevub}, dominated by the uncertainty on $m_b$. 

In practice the measurable rate is strongly reduced, since the background from
$B\to X_c\ell\nu$ decays (that dominates the signal by a factor $50$) must be suppressed 
by requiring stringent kinematic cuts. The reduced accessible rate break 
the convergence of the HQE and this increase considerably the theoretical uncertainty. 
The cuts usually relies on the fact that the $u$-quark mass is much lighter 
than the $c$-quark, as a consequence the distribution of some kinematics variables differs
between $b\to u$ and $b\to c$ transitions. For example the the lepton momentum $p_\ell$, 
extends to higher values for the signal, and analogously the distribution of the hadronic 
mass $m_X$ of the hadronic jet produced by the fragmentation of the quark $u$, 
extends toward lower values compared to the $m_X$ distribution for the quark $c$, that cannot 
be lower than the mass od the $D$ meson. It is so
possible to select regions of the phase space where the signal over background is 
reasonable, but usually the acceptances tend to be small (from $\circ 6\%$ requiring $p_\ell$ 
above the kinematic endpoint of the leptons from $B\to X_c\ell\nu$, 
to $\circ 70\%$ requiring $m_X$ lower than the mass of the $D$ meson) and 
finite experimental resolution have to be accounted. 

In the reduced phase space, the leading term of the non-perturbative 
correction to the expected rate, becomes of the order $\Lambda_{QCD}/m_b$ instead of 
$\Lambda^2_{QCD}/m^2_b$, and is described by the distribution function 
(called {\it shape function}, SF) of the momentum of the $b$ quark inside the $B$ meson. 
The SF cannot be computed perturbatively and must be determined
experimentally. It is a function of $m_b$ and of other heavy quark parameters that
describe the internal structure of the $B$ meson, which can be determined from the  
fit to the moments of the inclusive semileptonic $B\to X_c\ell\nu$ decays described above, 
and also directly from the $B\to X_s\gamma$ decays. 

Because the uncertainty on the SF parameters (including also $m_b$) 
is one of the biggest source of uncrtainty in the present $|V_{ub}|$ determinations, 
better understanding of the $B\to X_c\ell\nu$, 
$B\to X_s\gamma$ are important to reduce the uncertainty on $|V_{ub}|$. 
Additional uncertainty that require further theoretical studies are due to the sub-leading shape 
functions that affects in a different way the $b\to u$ transitions from the 
radiative or $b\to c$ transitions. 

Moreover in the limited region of the $B\to X_u\ell\nu$ phase space 
used to extract the signal yields, further complications
arises due to possible Weak Annihilation (WA) contributions, that affect differently $B^0$ from $B^+$ 
and other non-perturbative effects that contribute to the tail of the $B\to X_u\ell\nu$ phase space.
Experimental constraints to the WA will be discussed in the Sec.\ref{wa_section}.

The measurement of the partial branching ratio $\Delta {\cal B} (B\to X_u\ell\nu)$ 
can be translated into $|V_{ub}|$ by 
$|V_{ub}|=\sqrt{{\Delta \cal B}/{(\tau_B\cdot \Gamma_{th})}}$,
where $\tau_B$ is the $B$ average meson lifetime, and $\Gamma_{th}$ is the reduced decay rate
defined as $\Gamma_{th}=\Delta \Gamma_{th}/|V_{ub}|^2$, where $\Delta \Gamma_{th}$
is the partial width into the phase space defined by the kinematic cuts, predicted by
the theory. There are various calculations of the $\Delta \Gamma_{th}$ available, documented in
Refs.\cite{BLNP} (BLNP), \cite{GGOU} (GGOU),\cite{DGE} (DGE),\cite{ADFR} (ADFR), and \cite{BLL} (BLL).

%At the modern $B-$Factories, it is has been widely used the $B_{reco}$ sample described above, allowing 
%to perform analysis with large signal acceptance and reasonable S/N ratio. 

\begin{figure}[h]
\centering
\includegraphics[width=50mm]{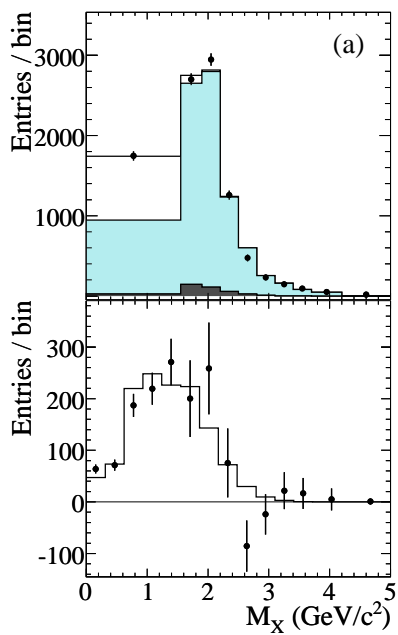}
\caption{Upper: Measured $M_x$ spectra. The result of the fit 
is superimposed: $B\to X_u\ell\nu$ signal generated inside (outside) 
the selected kinematic region $M_x<1.55$~GeV, in white histogram 
(grey histogram), and the $B\to X_c\ell\nu$ background (cyan).
Lower: background subtracted plot (no unfolded), with finer binning.}
\label{f:babar_vub}
\end{figure}

{\sc BaBar} recently published \cite{vub_babar}, using a sample of $382M$ of $B\overline B$ mesons, 
analyses on various kinematic variables, studying semileptonic decays with $p_\ell>1$~GeV, 
recoiling against a fully reconstructed $B$ mesons. 
The measured partial branching ratio have been obtained with cuts on 
the hadronic mass $m_X$ of the hadronic system $X_u$, 
the hadronic light-cone momentum $P_+=E_X-|p_x|$, and a combination of a 
cut on $m_X$ with a cut on the the invarian mass of the leptonic system $q^2=(p_\ell+p_\nu)^2$. 
In Fig.\ref{f:babar_vub}, as an example, is shown the distribution of the measured $m_X$ variables.
Within the signal region used to extract the partial branching ratio, $m_X<1.55$~GeV, an yield
of $803\pm60$ $Xu\ell\nu$ events has been extracted. 
The experimental systematics have large contributions from the modeling of the signal 
and $B\to X_c\ell\nu$ decays, and also from the detector efficiency to charged, photons and 
neutral hadrons.
The {\sc BaBar} results are reported in table \ref{t:partialrate}, together with the analogous 
partial branching ratio measured by Belle using a similar technique \cite{vub_belle}.
\begin{table}[h]
\begin{center}
\caption{Summary of the measured partial branching ratio $\Delta B$ in units of $10^{-3}$. 
The uncertainty are statistical and systematics (this include the signal model systematics). 
For Belle, the published $\Delta\Gamma(X_u\ell\nu)$ has been translated into a partial branching
ratio using the $B^{+/0}$ average lifetime from Ref.\cite{PDG06}.}
\begin{tabular}{l|c|c}
\hline \textbf{cut}   & \textbf{BaBar} \cite{vub_babar} & \textbf{Belle}\cite{vub_belle} \\
\hline 
$M_x<1.55$~GeV & 1.18$\pm$0.09$\pm$0.08       & - \\
\hline$M_x<1.70$~GeV & -                            & 1.21$\pm$0.11$\pm$0.09 \\
\hline$P_+<0.66$~GeV & 0.95$\pm$0.10$\pm$0.08       & 1.08$\pm$0.10$\pm$0.10 \\
\hline$M_x<1.7$~GeV    & 0.81$\pm$0.08$\pm$0.07 & 0.82$\pm$0.08$\pm$0.07 \\
$\&$ $q^2>8$~GeV$^2$ &  & \\
\hline
\end{tabular}
\label{t:partialrate}
\end{center}
\end{table}
The {\sc Babar} and Belle partial branching ratios, are in good agreements. But taking into 
account the large correlations between the analysis with the various kinematics cuts, both 
{\sc BaBar} and Belle observe a discrepancy at the level of $\approx$2.0-2.5$\sigma$ 
between the ratio of the measured $P_+<0.66$~GeV partial rate and the measured $m_X-q^2$
partial rate, compared with the corresponding ratio predicted by various theoretical calculations. 
This observation requires more investigations from both experimental and theoretical side, because 
can be an hint of the contributions of WA, or of other effects not accounted 
in the present theoretical calculations. 

\subsection{Determination of $|V_{ub}|$}

As we have said, the determination of $|V_{ub}|$ from the partial rates, require inputs
from theory. HFAG extract the averages between the various experiments and 
kineamtical variables, using all available calculations. In Tab.\ref{t:vub}
we report only results using the BLNP, GGOU and BLL calculations.
 
For the analysis in Ref.\cite{vub_babar} and Ref.\cite{vub_belle}, only 
the results for $m_X$ has been used, due to the large correlations with the $P_+$ and $q^2$
kinematic cuts. BLNP and GGOU agree pretty well, but the $m_b$ value used in input are different. 
If the same $m_b$ is used, than BLNP is more than $5\%$ higher than the GGOU average.
The BLL which is applied only for some cuts, give a value quite higher if compared
with the other methods. 
For a reliable comparison between different calculations, consistent set of input parameters
should be used, we hope in the near future, these inconsistecies will be fixed.

The sensitivity to the SF and to the WA can be strongly reduced if larger portion of 
the $B\to X_u\ell\nu$ phase space are integrated. An analysis performed by {\sc BaBar}
on a sample of $80$~fb$^{-1}$ \cite{babarnoshape}, integrates about $96\%$ of the total rate
is interesting for the smallness of the theoretical error (less than $3\%$), 
but the statistical uncertainty of $18\%$ reduce the impact of this measurement 
on the global averages.

At CKM2008, very recently, Belle presented \cite{phil} a preliminary analysis 
 based on a multivariate technique to reduce the large 
$B\to X_c\ell\nu$ background,integrates  about $90\%$ of the phase space.
This measurement extracts $|V_{ub}|$ with a total uncertainty of $7\%$, 
where the contribution due to the theory and $m_b$ is only $4\%$. More measurements of this type 
should be performed in the future exploiting the large dataset available at the B-Factoris.

\begin{table}[h]
\begin{center}
\caption{Inclusive determinations of $|V_{ub}|$ used by HFAG to perform the world
averages. 
The results of the analysis of $P_+$ and the combined $M_x-q^2$ from Refs.\cite{vub_babar} and
\cite{vub_belle},
are not included in the average because stongly correlated with the $M_x$ analyses. We report only 
the results with BLNP and GGOU and BLL. The input values ($m_b$ and $\mu_\pi^2$) are different: 
BLNP uses the results of the global fits with the exclusion of the $B\to X_s\gamma$ moments, instead
GGOU includes also the radiative moments.}
\begin{tabular}{l|c|c|c}
\hline \textbf{cut} & \textbf{BLNP} & \textbf{GGOU} & \textbf{BLL}\\
\hline 
$E_e$  \cite{cleo_endp}    & 3.52$\pm$0.41$^{+0.38}_{-0.32}$ &  3.70$\pm$0.43$^{+0.25}_{-0.39}$ & \\
$M_x,q^2$ \cite{belle_sim} & 3.98$\pm$0.42$^{+0.34}_{-0.28}$ &  4.15$\pm$0.44$^{+0.33}_{-0.34}$ & 4.71$\pm$0.50$^{+0.35}_{-0.35}$ \\
$E_e$  \cite{belle_endp}   & 4.36$\pm$0.41$^{+0.36}_{-0.30}$ &  4.55$\pm$0.42$^{+0.22}_{-0.31}$ & \\
$E_e$  \cite{babar_endp}   & 3.90$\pm$0.22$^{+0.35}_{-0.30}$ &  4.07$\pm$0.23$^{+0.23}_{-0.33}$ & \\
$E_e,s^{m}$ \cite{babar_s}    & 3.95$\pm$0.27$^{+0.42}_{-0.36}$ & &   \\
$M_x$ \cite{vub_belle}    & 3.66$\pm$0.27$^{+0.29}_{-0.24}$ &  3.89$\pm$0.26$^{+0.19}_{-0.22}$ & \\
$M_x$ \cite{vub_babar}    & 3.73$\pm$0.24$^{+0.33}_{-0.28}$ &  4.01$\pm$0.19$^{+0.26}_{-0.29}$ & \\
$M_x,q^2$\cite{vub_belle} & - & -  & 5.01$\pm$0.39$^{+0.37}_{-0.37} $\\
$M_x,q^2$ \cite{vub_babar}& - & -  & 4.92$\pm$0.32$^{+0.36}_{-0.36} $\\
\hline
Average   & 3.98$\pm$0.14$^{+0.32}_{-0.27}$ &  3.94$\pm$0.15$^{+0.20}_{-0.23}$ & 4.91$\pm$0.24$^{+0.38}_{-0.38} $\\
\hline
\end{tabular}
\label{t:vub}
\end{center}
\end{table}

\subsection{$M_X$ hadronic moments in $B\to X_u \ell\nu$ decays} 

\begin{figure}[h]
\centering
\includegraphics[width=65mm]{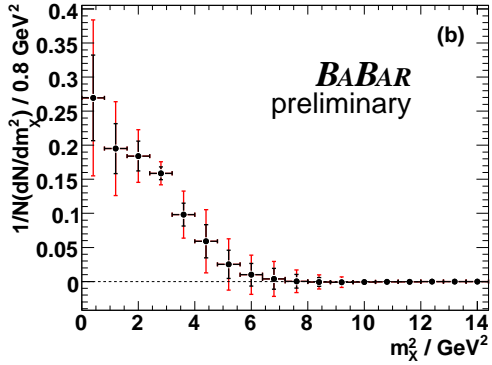}
\caption{Unfolded hadronic mass spectrum in $B\to X_u\ell\nu$. The inner error
bars show the statistical uncertainty only.} \label{}
\label{f:kerstin}
\end{figure}

The determination of the OPE parameters from the measurements of the hadronic moments in
charmless $B\to X_u\ell\nu$ decays, is important to test the theoretical framework used to
extract $|V_{ub}|$. The measurement of the $X_u$ hadronic moments has been performed for 
the first time by {\sc BaBar} \cite{kerstin}. The analysis is based on the sample and same 
technique of the analysis used to extract $|V_{ub}|$. The $B\to X_c\ell\nu$ background is subtracted
by a fit to the hadronic mass spectrum, the result of the fit is reported in Fig.\ref{f:kerstin}.
The $m_X^2$ spectrum is unfolded for the detector acceptance, efficiency and resolution effects
and the first, second and third central moments are extracted from the unfolded spectrum. 
An  HQE fit of these moments, in the kinetic scheme \cite{kinetic} yields
\begin{eqnarray}
m_b      & = & (4.604\pm 0.125 \pm 0.193\pm 0.097) {\rm GeV}\nonumber \\
\mu_\pi^2& = &(0.398 \pm 0.135 \pm 0.195\pm 0.036) {\rm GeV}^2 \nonumber
\label{babar_rismom}
\end{eqnarray}
\noindent where the first error is statistical, the second 
is systematics and the thirds comes from the theory. 
The results for $m_b$ and $\mu_\pi^2$ are consistent within the reported uncertainty, 
with the determinations with $B\to X_c\ell \nu$ and $B\to X_s\gamma$ events. Despite the large
uncertainty, this studies deserve more studies in future.

\subsection{Weak annihilation}
\label{wa_section}

One of the effects that is not included in current calculations 
of the partial decay rate, is the WA \cite{bigi_wa}, which is expected to 
contribute at the level of a few percent \cite{neubert_wa,voloshin_wa,ligeti_wa,gambino_wa}. 
Simply speaking, WA refers to the annihilation 
of the $b-{\bar u}$ pair to a virtual $W$ boson, and results in an enhancement 
of the decay rate near the endpoint of the $q^2$ spectrum. Here $q^2$ refers to 
the mass squared of the virtual $W$.

Experimentally, WA should be observable as a violation of isospin invariance, 
i.e. difference in the partial decay rates of $B^0 \to X_{u}^- \ell^+ \nu$ and 
$B^+ \to X_{u}^0 \ell^+ \nu$, at high $q^2$, since it occurs only for charged $B$ mesons. 

{\sc BaBar} performed a first measurement of the partial branching fraction for 
inclusive $B^0\to X_{u}^- \ell^+ \nu$ decays 
above 2.3~GeV$/c$ of the charged lepton momentum \cite{babar:wa}. $B^0{\bar{B}}^0$ events 
produced at the $\Upsilon(4S)$ resonance are tagged by the partially reconstructed  
$B^0\to D^{*+}\ell\nu$ decays via the reconstruction of the soft pion $\pi_{soft}^+$
emitted by the $D^{*+}$ decays. The signal is extracted fitting the distribution of
the neutrino mass of the tag side $M_\nu^2$, that is peaked at $0$ for true 
$B\to\ell$ events. The large $B\to X_c\ell\nu$ background is suppressed 
using the kaon veto and the soft pion veto on the signal side. The remaining 
charm background is fixed to the Monte Carlo estimation, and the large combinatoric 
background, is modelled using the wrong charge sample in which the 
soft pion and the lepton on the tag side, have the same charge.

We identify the charmless semileptonic decay of 
the second $B$ meson in the event and compare its partial decay rate with 
the partial rate for the sum of charged and neutral $B$ mesons previously 
published \cite{bad1047}, and extract the difference in these partial decay 
rates between $B^+$ and $B^0$ mesons. 
\begin{figure}
\includegraphics[width=8cm]{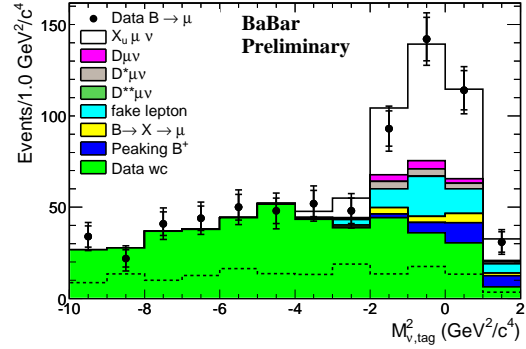}\\ 
\caption{ $M_\nu^2$ distribution requiring $2.3<P_{\ell}<2.6$~GeV/$c$ for $\mu$ sample. 
The signal component from simulation and the wrong-charge sample 
have been rescaled according to the fit results. The inner error bars are the
statistical error from the right-charge sample only while the larger error bars 
include also the statistical errors of the 
wrong-charge sample and of the various peaking components described by the simulation.}
\label{f:mnufit}
\end{figure}
The result for the partial branching fraction for the interval
$2.3 < p_\ell <2.6$~Gev is $\Delta{\cal B}(B^0\to X_u\ell\nu)=
(1.30\pm0.21\pm0.07)\times 10^{-4}$. Combining this result with the
inclusive lepton spectra allows to extract a ratio of the $\Gamma(B^0)$ and
$\Gamma(B^+)$ decays widths for charmless semileptonic decays,
 $R^{+/0}= \frac{\Delta \Gamma^+}{\Delta \Gamma^0}=1.18 \pm 0.35 \pm 0.17$.
Thus with the presently available data sample, there is no evidence for a difference 
in partial decay rates in $B^0$ and $B^+$ at the high end of the lepton momentum spectrum, 
where we would expect the impact of WA in $B^+$ decays. Defining 
 $\Delta\Gamma_{WA}=\Delta \Gamma^+-\Delta \Gamma^0$ the contribution of the WA, the following limit 
ca be set
\begin{equation} 
\frac{|\Gamma_{WA}|}{\Gamma_{u}} < \frac{3.8~\%}{f_{WA}(2.3-2.6)},{~~~\rm~at~90\% ~C.L.}~, \nonumber\\
\end{equation}
\noindent where $f_{WA}(2.3-2.6)$ refers to the fraction of the 
weak annihilation rate contributing in the momentum interval $(2.3-2.6)$~GeV$/c$.
This limit is also consistent with a model dependent limit set by CLEO 
\cite{cleo:wa} studing the $q^2$ spectra.  

These limits can be translated in an uncertainty on $|V_{ub}|$ of the
order of $2-3\%$ according to the acceptance of the kinematic cut. 
The flavor dependent partial decay rate could also be determination, using other
$B-flavor$ tagging techniques, like the fully reconstructed hadronic sample
described above. These studies should be performed in the future.

\section{Conclusions}
The CKM matrix elements $V_{cb}$ and $V_{ub}$ are fundamental parameters of the Standard Model.
$|V_{cb}|$ with the inclusive decays already reached an error well below the $\approx 2\%$.
Future progress from the experimental side are important, in particular it is crucial to 
understand the model of the $X_c$ state which uncertainty is now dominated by the 
scarce knowledge of the $B\to D^{**}\ell\nu$ rates. At present the inclusive determination is only
marginally compatible with the exclusive determination. The two determinations differs by
$\approx 2\sigma$. These discrepancy have to be understood in the next future.
The total uncertainty on $V_{ub}$ is below $10\%$. It is crucial to reduce this uncertainty
for precise testing of the CKM picture of $CP$ violation and for indirect searches of New Physics.
There are many theoretical framework available on the market. The $B-$factories have 
enough statistics that detailed studies of the $B\to X_u\ell\nu$ kinematic can be exploited
to prune the different available calculations. {\sc Babar} moved forward in this direction
starting to study the moments of the hadronic mass in semileptonic $B\to X_u$ transitions,
and the rate of $B\to X_u\ell\nu$ decays, in $B^0$ tagged events. 

%%%%%%%%%%%%%%%%%%%%%%%%%%%%%%%%%%%%%%%%%%%%%%%%%%%%%%%%%%%%%%%%%%%%%%%%%%%%%%%%%%%%%%%%

\bigskip 

%\begin{thebibliography}{9}   % Use for  1-9  references
%
%\end{thebibliography}


\begin{thebibliography}{99} % Use for 10-99 references

\bibitem{beta} B.Aubert {\it et al.}, Phys.Rev.Lett {\bf 94}, 161803 (2005); K.Abe {\it et al.}, 
              Phys.Rev.Lett {\bf 87}, 091802 (2005); 

\bibitem{CKM} N. Cabibbo, Phys. Rev. Lett. {\bf 10}, 531 (1963).
M. Kobayashi and T. Maskawa, Prog. Theor. Phys. {\bf 49}, 652 (1973).

\bibitem{ope} A.V.Manoher and M.B.Wise, Phys. Rev. D {\bf 49}, 1310 (1994);
              I.I.Bigi {et al.}, Phys. Rev. Lett. {\bf 71}, 496 (1993);
              I.I.Bigi {et al.}, Phys. Lett. {\bf B323}, 408 (1994).
% VCB
\bibitem{CLEO_mom} N.E.Adams {\em et al.} (CLEO Collab.), Phys. Rev. {\bf D70}, 032002 (2007);
A.H.Mahmood {\em et al.} (CLEO Collab.), Phys. Rev. {\bf D70}, 032003 (2007).
\bibitem{BaBar_mom} B.Aubert {\em et al.} ({\sc BaBar} Collab.), Phys. Rev. {\bf D69}, 111103 (2004);
 B.Aubert {\em et al.} ({\sc BaBar} Collab.), Phys. Rev. {\bf D69}, 111104 (2004).
\bibitem{Belle_mom} C.Schwanda {\em et al.} (Belle Collab.), Phys. Rev. {\bf D75}, 032005 (2007);
P.Urquijo {\em et al.} (Belle Collab.), Phys. Rev. {\bf D75}, 032001 (2007).
\bibitem{DELPHI_mom} J.Abdallah {\em et al.} (DELPHI Collab.), Eur. Phys. J {\bf C45}, 35 (2006).
\bibitem{CDF_mom} D.Costa {\em et al.} (CDF Collab.), Phys. Rev. {\bf D71}, 051103 (2005).

\bibitem{BaBar_newmom} B.Aubert {\em et al.} ({\sc BaBar} Collab.), arXiv:0707.2670.
\bibitem{gambino_mixmom} P. Gambino {\em et al.}, JHEP {\bf 09}, 010, (2005).
\bibitem{BaBar_gamma} B.Aubert {\em et al.} ({\sc BaBar} Collab.), Phys. Rev. {\bf D72}, 052004 (2005).
\bibitem{kinetic} I.I.Y.Bigi {\em et al.} Phys. Rev {\bf D56}, 4017, (1997);
I.I.Y.Bigi {\em et al.} Phys. Rev {\bf D52}, 196, (1995);
P. Gambino and N. Uraltsev, Eur. Phys. J. {\bf C34}, 181, (2004).
\bibitem{Belle_egamma} P.Koppenburg {\em et al.} (Belle Collab.), Phys. Rev. Lett. {\bf 93}, 061803 (2004).
\bibitem{Belle_newegamma} C.Schwanda {\em et al.} (Belle Collab.), arXiv:0803.2158.
\bibitem{CLEO_egamma} S.Chen {\em et al.} (CLEO Collab.), Phys. Rev. Lett. {\bf 87}, 251807 (2001).
\bibitem{1s} C.W.Bauer {\em et al.}, Phys. Rev. {\bf D70}, 094017 (2004).
\bibitem{HFAG} See http://www.slac.stanford.edu/xorg/hfag/\\semi/pdg08/home.shtml

\bibitem{Neubert_lp07} M. Neubert plenary talk at Lepton-Photon 2007.
\bibitem{laiho} C.Bernard {\em et al.}, arXiv:0808.2519, submitted to Phys.Rev.D

% VUB
\bibitem{opevub} N.Uraltsev, Int. J. Mod. Phys. A {\bf 14}, 4641 (1999).
\bibitem{BLNP} B.O.Lange, M.Neubert,G.Paz, Phys. Rev. D {\bf 72}, 073006 (2005). 
\bibitem{GGOU} P. Gambino {\it et al}, JHEP {\bf 0710}, 058 (2007).
\bibitem{DGE} J.R.Andersen and E.Gardi, JHEP {\bf 0601}, 097 (2006). 
\bibitem{ADFR} U. Aglietti {\it et al}, arXiv:0711.0860.
\bibitem{BLL}  C.W.Bauer {\it et al.}, Phys. Rev. D {\bf 064}, 113004 (2001). 
\bibitem{vub_babar} B.Aubert {\it et al.}, ({\sc BaBar} Coll.) Phys. Rev. Lett. {\bf 100}, 171802 (2008). 
\bibitem{PDG06} W. M. Yao {\it et al.}, Journal of Physics {\bf G33}, 1, (2006).
\bibitem{vub_belle} Bizjak {\it et al.}, (Belle Coll.) Phys. Rev. Lett. {\bf 95}, 2418012 (2005). 
\bibitem{cleo_endp} A. Bornheim {\it et al.}, (CLEO Coll.) Phys. Rev. Lett. {\bf 88}, 231803 (2002).
\bibitem{belle_sim} H. Kakuno {\it et al.}, (Belle Coll.) Phys. Rev. Lett. {\bf 92}, 101801 (2004).
\bibitem{belle_endp} A. Limosani {\it et al.}, (Belle Coll.) Phys. Lett. B {\bf 621}, 28 (2005).
\bibitem{babar_endp} B.Aubert {\it et al.}, ({\sc BaBar} Coll.) Phys. Rev. D {\bf 73}, 012006 (2006).
\bibitem{babar_s} B.Aubert {\it et al.}, ({\sc BaBar} Coll.) Phys. Rev. Lett {\bf 95}, 111801 (2005).
%
\bibitem{babarend} B.Aubert {\it et al.}, Phys. Rev. D {\bf 73}, 012006 (2006) 
\bibitem{LLR} A.K.Leibovich, I.Low and I.Z.Rothstein, Phys. Rev. D {\bf 61}, 053006 (2000) 
\bibitem{bsg} B.Aubert {\it et al.}, Phys. Rev. D {\bf 72}, 052004 (2005) 
\bibitem{babarnoshape} B.Aubert {\it et al.}, Phys. Rev. Lett {\bf 96}, 221801 (2006).
\bibitem{phil} P.Urquijio, presentation at CKM 2008, Rome, 9-13 Sept, 2008 (http://ckm2008.roma1.infn.it/).
%\bibitem{hfag} Heavy Flavor Averaging Group, hep-ex/0603003 
%\bibitem{CKM} N. Cabibbo, Phys. Rev. Lett. {\bf 10}, 531 (1963).
%M. Kobayashi and T. Maskawa, Prog. Theor. Phys. {\bf 49}, 652 (1973).
\bibitem{PDG_bob} See the review on $V_{ub}$ and $V_{cb}$ available in \cite{PDG06}
\bibitem{kerstin} K.Tackmann (on behalf of {\sc BaBar} Coll.), SLAC-PUB-13036, arXiv:0801.2985 [hep-ex].
\bibitem{bigi_wa} I.I.Bigi and N.G.Uraltsev, Nucl. Phys. {\bf B423}, 33 (1994) .
\bibitem{neubert_wa} M. Neubert and C. T. Sachrajda, Nucl. Phys. {\bf B483}, 3339, (1997).
\bibitem{voloshin_wa} M.B.Voloshin, Phys. Lett. {\bf B515}, 74, (2001).
\bibitem{ligeti_wa} A.K.Leibovich, Z.Ligeti and M.B.Wise, Phys. Lett. {\bf B539}, 242, (2002).
\bibitem{gambino_wa} P. Gambino, G. Ossola, and N. Uraltsev, JHEP {\bf 09}, 010, (2005).
\bibitem{babar:wa} {\sc BaBar} Collaboration (B. Aubert et al.), SLAC-PUB-12740, arXiv:0708.1753 [hep-ex].
\bibitem{bad1047} {\sc BaBar} Collaboration (B. Aubert et al.), Phys. Rev. {\bf D73}, 012006, (2006). 
\bibitem{partial} This technique was originally applied to $B\to D^*\ell\nu$ decays by ARGUS: 
ARGUS Collaboration, H Albrecht {\em at al.} Phys. Lett. {\bf B324}, 249 (1994).

\bibitem{cleo:wa} CLEO Collaboration, J. L. Rosner {\em et al.}, Phys. Rev. Lett {\bf 96}:121801 (2006). 

\end{thebibliography}
\end{document}